\documentclass[letterpaper]{jpconf}
\usepackage{graphicx}
\usepackage{cite}
\begin{document}
\title{%
$\;$\,\\[-7ex]
\hspace*{\fill}{\tt\normalsize ANL-PHY-11081-TH-2004}\\[1ex] %
Pseudoscalar and vector mesons as $q\bar{q}$ bound states}

\author{A.~Krassnigg$^1$ and P.~Maris$^2$}

\address{$^1$ Physics Division, Argonne National Laboratory,
Argonne, IL 60439}
\address{$^2$ Department of Physics and Astronomy, University of Pittsburgh,
Pittsburgh, PA 15260}

\ead{andreas.krassnigg@anl.gov, pim6@pitt.edu}

\begin{abstract}
Two-body bound states such as mesons are described by solutions of the
Bethe--Salpeter equation.  We discuss recent results for the
pseudoscalar and vector meson masses and leptonic decay constants,
ranging from pions up to $c\bar{c}$ bound states. Our results are in
good agreement with data.  Essential in these calculation is a
momentum-dependent quark mass function, which evolves from a
constituent-quark mass in the infrared region to a current-quark mass
in the perturbative region.  In addition to the mass spectrum, we
review the electromagnetic form factors of the light mesons.
Electromagnetic current conservation is manifest and the influence of
intermediate vector mesons is incorporated self-consistently.  The
results for the pion form factor are in excellent agreement with
experiment.
\end{abstract}

\section{Dyson--Schwinger equations}
The set of Dyson--Schwinger equations form a Poincar\'{e} covariant
framework within which to study hadrons~\cite{review,Maris:2003vk}.
In rainbow-ladder truncation, they have been successfully applied to
calculate a range of properties of the light pseudoscalar and vector
mesons, see Ref.~\cite{Maris:2003vk} and references therein.  

The DSE for the renormalized quark propagator $S(p)$ in Euclidean
space is~\cite{review}
\begin{equation}
\label{genDSE}
 S(p)^{-1} = i \, Z_2(\zeta) \,/\!\!\!p + Z_4(\zeta)\,m(\zeta) +
        Z_1(\zeta) \int\!\!\frac{d^4q}{(2\pi)^4} \,
    g^2 D_{\mu\nu}(p-q) \, \textstyle{\frac{\lambda^i}{2}}
    \gamma_\mu \, S(q) \, \Gamma^i_\nu(q,p) \;,
\end{equation}
where $D_{\mu\nu}(p-q)$ and $\Gamma^i_\nu(q;p)$ are the renormalized
dressed gluon propagator and quark-gluon vertex, respectively.  
The most general solution of Eq.~(\ref{genDSE}) has the form
\mbox{$S(p)^{-1} = i \,/\!\!\! p A(p^2) + B(p^2)$}, renormalized at
spacelike $\zeta^2$ according to \mbox{$A(\zeta^2)=1$} and
\mbox{$B(\zeta^2)=m(\zeta)$} with $m(\zeta)$ the current quark mass.

Mesons are described by solutions of the homogeneous Bethe--Salpeter
equation (BSE)
\begin{equation}
 \Gamma_H(p_+,p_-;P) = \int\!\!\frac{d^4q}{(2\pi)^4} \,
        K(p,q;P) \; S(q_+) \, \Gamma_H(q_+,q_-;P) \, S(q_-)\, ,
\label{homBSE}
\end{equation}
at discrete values of $P^2 = -M_H^2$, where $M_H$ is the meson mass.
In this equation, $p_+ = p + P/2$ and $p_- = p - P/2$ are the outgoing
and incoming quark momenta respectively, and similarly for $q_\pm$.
The kernel $K$ is the renormalized, amputated $q\bar q$ scattering
kernel that is irreducible with respect to a pair of $q\bar q$ lines.
Together with the canonical normalization condition for $q\bar q$
bound states, Eq.~(\ref{homBSE}) completely determines the bound state
Bethe--Salpeter amplitude (BSA) $\Gamma_H$.  Different types of
mesons, such as pseudoscalar or vector mesons, are characterized by
different Dirac structures.

To solve the BSE, we use the ladder truncation,
\begin{equation}
        K(p,q;P) \to
        - 4\pi \alpha^{\rm eff}\big((p-q)^2\big) \,
    D^0_{\mu\nu}(p-q)
        \textstyle{\frac{\lambda^i}{2}}\gamma_\mu \otimes
        \textstyle{\frac{\lambda^i}{2}}\gamma_\nu \, ,
\end{equation}
in conjunction with the rainbow truncation for the quark DSE,
Eq.~(\ref{genDSE}): \mbox{$\Gamma^i_\nu(q,p) \rightarrow
\frac{\lambda^i}{2}\gamma_\nu$} and \mbox{$Z_1 g^2 D_{\mu \nu}(k)
\rightarrow 4\pi\alpha^{\rm eff}(k^2) D^0_{\mu\nu}(k)$}, with $k=p-q$.
Here, $D^0_{\mu\nu}(k)$ is the free gluon propagator in Landau gauge,
and $\alpha^{\rm eff}(k^2)$ an effective quark-quark interaction,
which reduces to the one-loop running coupling of perturbative QCD
(pQCD) in the perturbative region.  This truncation preserves both the
vector Ward--Takahashi identity (WTI) for the $q\bar q\gamma$ vertex
and the axial-vector WTI, independent of the details of the effective
interaction.  The latter ensures the existence of massless
pseudoscalar mesons associated with dynamical chiral symmetry breaking
(D$\chi$SB) in the chiral limit~\cite{Maris:1997tm,Maris:1997hd}.  In
combination with an impulse approximation, the former ensures
electromagnetic current conservation~\cite{Roberts:1994hh,emf}.

\section{Quark propagator: an overview}
A momentum-dependent quark mass function $M(p^2) = B(p^2)/A(p^2)$ is
central to QCD.  In the perturbative region this mass function
gives the one-loop perturbative running quark mass
\begin{eqnarray}
  M(p^2) &\simeq& \frac{\hat{m}}
  {\left(\case{1}{2}\ln\left[p^2
   /\Lambda_{\rm QCD}^2\right]\right)^{\gamma_m}} \,,
\label{eq:currentM}
\end{eqnarray}
with the anomalous mass dimension $\gamma_m = 12/(33-2N_f)$.
Dynamical chiral symmetry breaking means that this mass function is
nonzero even though the current-quark masses are zero.  In the chiral
limit the mass function is~\cite{Politzer:1976tv}
\begin{eqnarray}
 M_{\hbox{\scriptsize{chiral}}}(p^2) & \simeq & \frac{2\pi^2\gamma_m}{3}\,
        \frac{-\,\langle \bar q q \rangle^0}{p^2 \left(
        \frac{1}{2}\ln\left[p^2/\Lambda_{\rm QCD}^2\right]
    \right)^{1-\gamma_m}}\,,
\label{eq:chiralM}
\end{eqnarray}
with $\langle \bar q q \rangle^0$ the
renormalization-point-independent vacuum quark
condensate~\cite{Maris:1997tm}.

\begin{figure}[tb]
\includegraphics[width=18pc]{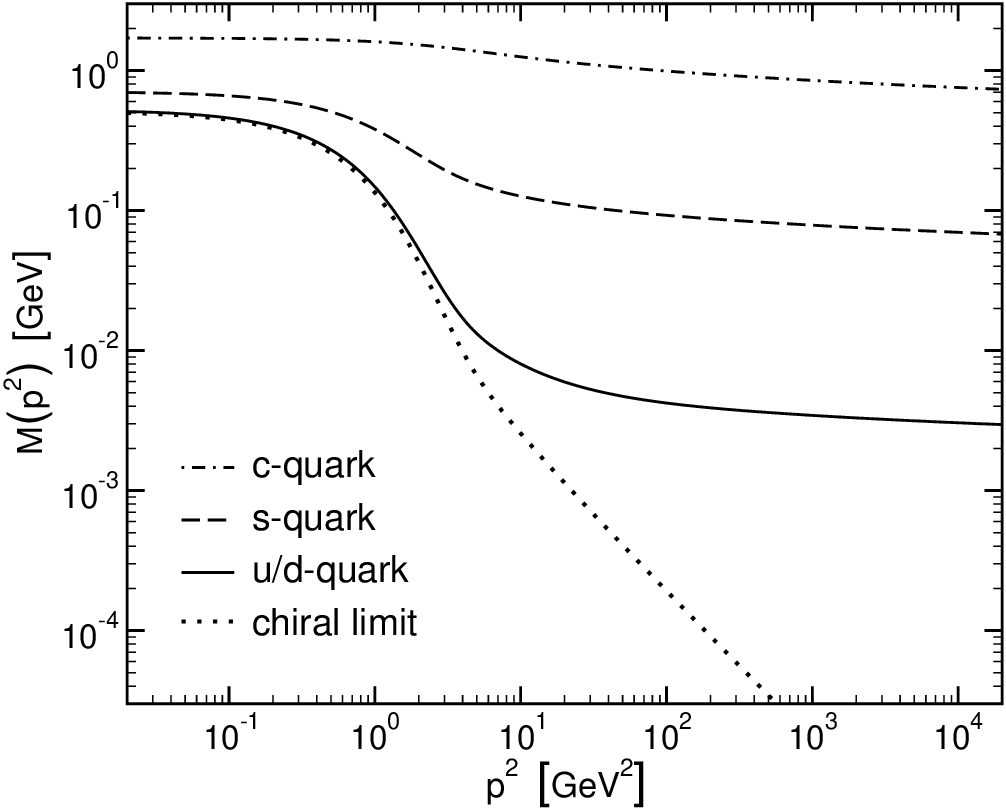}\hspace{1pc}%
\includegraphics[width=18pc]{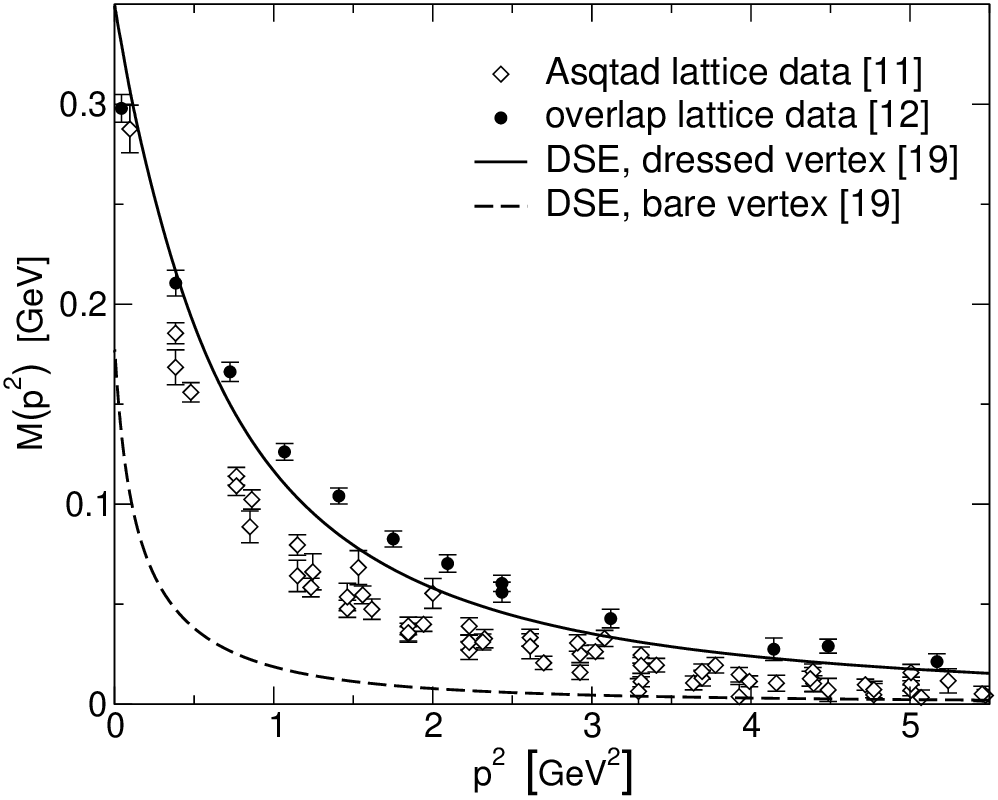}
\caption{\label{Fig:quark} 
The dynamical quark mass function $M(p^2)$ for different quark flavors
(left, adapted from Ref.~\cite{Maris:1997tm}) and a comparison of the
chiral-limit quark mass function with the lattice
data~\cite{Bowman:2002bm,Zhang:2003fa} (right, adapted from
Ref.~\cite{Fischer:2003rp}).}
\end{figure}

It is a longstanding prediction of DSE studies in QCD that the dressed
quark propagator receives strong momentum-dependent corrections at
infrared momenta, see e.g. Refs.~\cite{review,Maris:2003vk} and
references therein.  Provided that the (effective) quark-quark
interaction reduces to the perturbative running coupling in the
ultraviolet region, it is also straightforward to reproduce the
asymptotic behavior of Eqs.~(\ref{eq:currentM}) and
(\ref{eq:chiralM})~\cite{Higashijima:1983gx,Fomin:1984tv}.  Both these
phenomena are illustrated in the left panel of Fig.~\ref{Fig:quark}.
In this figure one can also see that the dynamical mass function of
the $u$ and $d$ quarks becomes very similar to that of quarks in the
chiral limit in the infrared region.  This is a direct consequence of
D$\chi$SB, and leads to a mass function of the order of several
hundred MeV for the light quarks in the infrared, providing one with a
constituent mass for quarks inside hadrons, even though the
corresponding current quark masses are only a few MeV.

These predictions were recently confirmed in lattice simulations of
QCD~\cite{lattice,Bowman:2002bm,Zhang:2003fa}.  Quantitative agreement
between the lattice simulations and the DSE results for the quark
propagator functions can be obtained within the rainbow truncation via
a suitable choice for the effective quark-quark
interaction~\cite{Maris:2002mt}.  Pointwise agreement for a range of
quark masses requires this interaction to be
flavor-dependent~\cite{Bhagwat:2003vw}, suggesting that dressing the
quark-gluon vertex $\Gamma^i_\nu(q,p)$ is important.  Indeed, both
lattice simulations~\cite{Skullerud:vertex} and DSE
studies~\cite{Bhagwat:2004hn,Bhagwat:2004kj,LlanesFischerAlkofer} of
this vertex indicate that $\Gamma^i_\nu(q,p)$ deviates significantly
from a bare vertex in the nonperturbative region.  A
(flavor-dependent) nonperturbative vertex dressing could make a
significant difference for the solution of the quark
DSE~\cite{Bhagwat:2004hn,Fischer:2003rp}, as can be seen from the
right panel of Fig.~\ref{Fig:quark}.  The consequences of a dressed
vertex for the meson BSEs are currently being
explored~\cite{Bhagwat:2004hn} and indications are that in the
pseudoscalar and vector channels, the effects are
small~\cite{Bhagwat:2004hn,Bender:1996bb,Bender:2002as}.

\section{Pseudoscalar mesons: ground and excited states}
The meson spectrum contains three pseudoscalars with quantum numbers
$I^G (J^P) L = 1^- (0^-) S$ and masses below $2\,$GeV: $\pi(140)$;
$\pi(1300)$; and $\pi(1800)$.  In a constituent-quark model, these
mesons are viewed as the first three members of a $q\bar q$ $n\,
^1\!S_0$ trajectory, where $n$ is the principal quantum number with
the ground state $\pi_0$ (the $\pi(140)$), and the others are its
first two radial excitations, $\pi_1$ and $\pi_2$.  The pseudoscalar
trajectory is particularly interesting, because its lowest mass member
is QCD's Goldstone mode.  Therefore an explanation should describe
both chiral symmetry and its dynamical breaking as well as a
correlation of the ground and excited states via an approximately
linear radial Regge trajectory.  The latter is easily realized in
Poincar\'{e} invariant quantum mechanics~\cite{Krassnigg:2003gh} but
the former is not.

\subsection{Chiral symmetry}
The chiral properties of QCD are manifest in the axial-vector WTI,
which reads
\begin{eqnarray}
P_\mu \Gamma_{5\mu}(q_+,q_-;P) &=&
 S^{-1}(q_+) i \gamma_5 + i \gamma_5 S^{-1}(q_-)
 - \, 2i\,m_q(\zeta) \,\Gamma_5(q_+,q_-;P) \, ,
\label{eq:avwti}
\end{eqnarray}
with $\Gamma_{5\mu}(q_+,q_-;P)$ and $\Gamma_5(q_+,q_-;P)$ the
renormalized dressed axial-vector and pseudoscalar vertices, each
satisfying an inhomogeneous extension of Eq.~(\ref{homBSE}).
Equation~(\ref{eq:avwti}) is an exact statement in QCD implying that
the kernels of the quark DSE, Eq.~(\ref{genDSE}), and of the BSEs have
to be intimately related.  A weak coupling expansion of the DSEs
yields perturbation theory and satisfies this constraint, but is not
useful in the study of intrinsically nonperturbative phenomena.
However, a systematic and symmetry preserving nonperturbative
truncation scheme
exists~\cite{Bhagwat:2004hn,Bender:1996bb,Bender:2002as}, allowing for
both elucidation and illustration of the consequences of the
axial-vector WTI.

Pseudoscalar mesons appear as pole contributions to the axial-vector
and pseudoscalar vertices at $P^2 = -M_{\pi_n}^2$.  The residues of
these poles are
\begin{eqnarray}
 f_{\pi_n} \, P_\mu &=&
 Z_2(\zeta)\, \int\!\!\frac{d^4q}{(2\pi)^4} \!
{\rm Tr}\left[ \gamma_5\gamma_\mu \;
        S(q_+) \, \Gamma_{\pi_n}(q_+,q_-;P) \, S(q_-)\, \right] \,,
\\
 i  \rho_{\pi_n}\!(\zeta) &=&
 Z_4(\zeta)\,\, \int\!\!\frac{d^4q}{(2\pi)^4}
{\rm Tr}\left[ \gamma_5 \;
        S(q_+) \, \Gamma_{\pi_n}(q_+,q_-;P) \, S(q_-)\, \right] \,,
\end{eqnarray}
both of which are gauge invariant and cutoff independent.  It follows
from Eq.~(\ref{eq:avwti}) that these residues satisfy the following
exact identity in QCD~\cite{Maris:1997hd,Krassnigg:2003wy}
\begin{eqnarray}
 f_{\pi_n} M_{\pi_n}^2 &=& 2 \, m(\zeta)  \,
 \rho_{\pi_n}(\zeta)\,,
\label{eq:gmorgen}
\end{eqnarray}
valid for every $0^-$ meson~\cite{Holl:2004fr}, irrespective of the
magnitude of the current quark mass~\cite{Ivanov:1998ms}.

For the $\pi_0$, D$\chi$SB yields: $\;\lim_{\hat m \to 0}\, f_{\pi_0}
\neq 0$ and $\;\lim_{\hat m \to 0}\,\rho(\zeta)= - 2 \langle \bar q q
\rangle^0_\zeta/f^0_{\pi_0}\neq 0$.  Hence, the
Gell-Mann--Oakes--Renner relation emerges as a corollary of
Eq.~(\ref{eq:gmorgen}) and the ground state pion is QCD's Goldstone
mode~\cite{Maris:1997hd}.  For the $n\geq 1$ pseudoscalar mesons one
has $M_{\pi_{n\geq 1}}>M_{\pi_0}$ by assumption, and hence $M_{\pi_{n>
0}} \neq 0$ in the chiral limit.  Furthermore, the ultraviolet
behavior of the quark-antiquark scattering kernel in QCD guarantees
that $\rho_{\pi_{n}}(\zeta)$ is finite in the chiral limit.  Hence, it
is a necessary consequence of chiral symmetry and the axial-vector WTI
that $f_{\pi_n}$ vanishes in the chiral limit for all excited
pseudoscalar mesons (i.\,e.~for all $n>0$)~\cite{Holl:2004fr}.

\subsection{Numerical results}
We now illustrate the exact results reviewed above in a model that
both preserves QCD's ultraviolet properties and exhibits D$\chi$SB,
namely the rainbow-ladder truncation of the set of DSEs.  For the
infrared behavior of the effective quark-quark interaction,
$\alpha^{\rm eff}(k^2)$, we employ an
Ansatz~\cite{Maris:1997tm,Maris:1999nt} that is sufficiently enhanced
in the infrared to produce a realistic value for the vacuum quark
condensate of about $(240\,{\rm GeV})^3$.  The model parameters, along
with the quark masses~\cite{Maris:1999nt,Hoell:2004un}, are fitted to
give a good description of the chiral condensate, $M_{\pi/K/\eta_c}$
and $f_{\pi}$.  The obtained quark propagator functions agree
qualitatively with lattice simulations.

\begin{figure}[t]
\includegraphics[width=18pc]{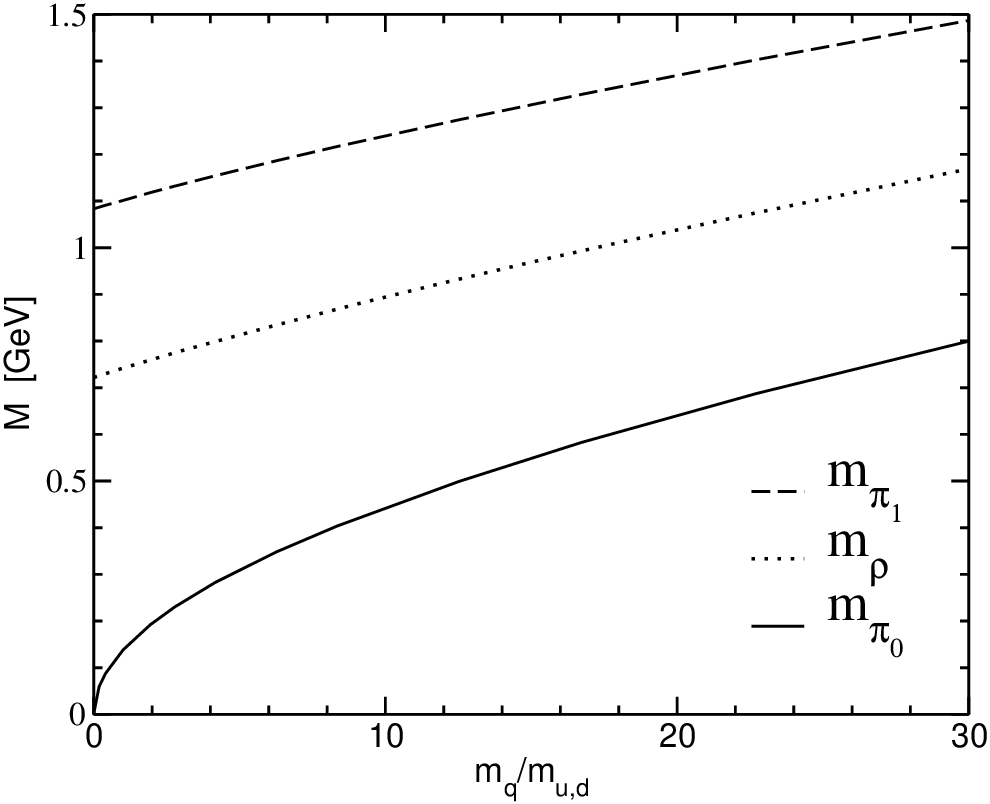}\hspace{1pc}%
\includegraphics[width=18pc]{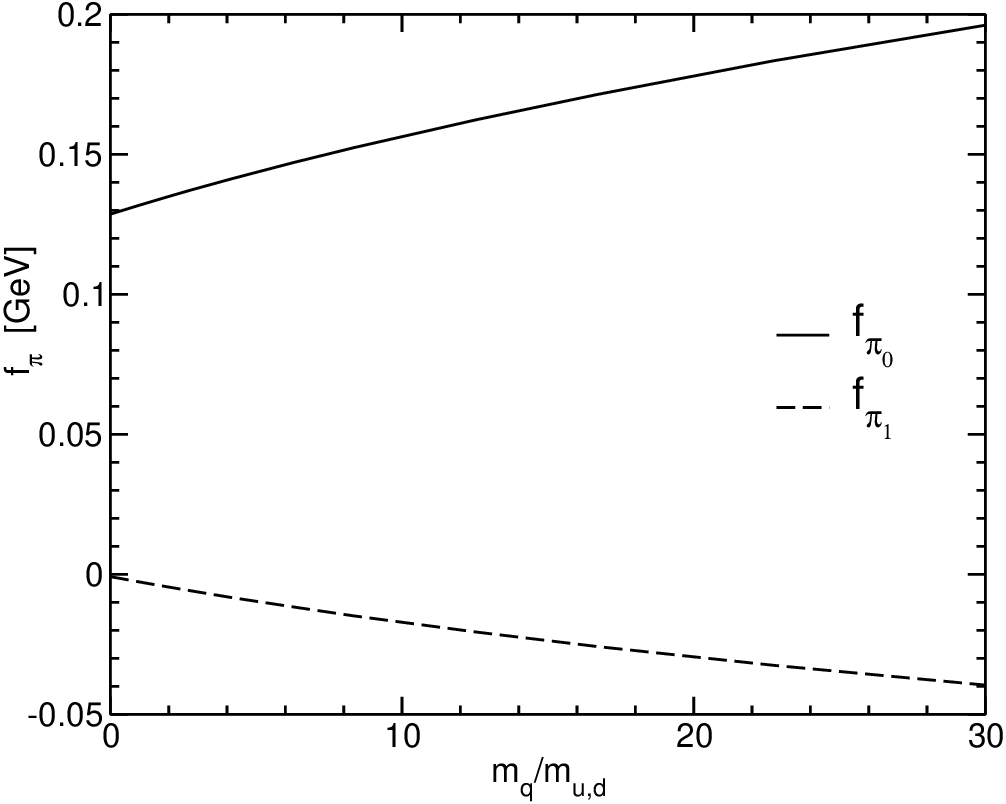}
\caption{\label{Fig:mesons}
The masses $M$ of the pseudoscalar ground and first radially excited
states (left) and the corresponding weak decay constants
(right) as functions of the current-quark mass $m_q$.  For comparison,
we also include the mass of the ground-state vector mesons.}
\end{figure}

\begin{table}[b]
\caption{\label{Tab:excpion} 
Results for the $q\bar{q}$ $0^{-}$ ground and first radially excited
states as well as $q\bar{q}$ $1^{-}$ ground states.  The current-quark
masses are $m_{u/d}=5.4\,{\rm MeV}$, $m_s=124\,{\rm MeV}$, and
$m_c=1.34\,{\rm GeV}$ at $\zeta=1\,$GeV, and all quantities are given in
GeV; for the leptonic decay constants we follow the conventions of
Ref.~\cite{Ivanov:1998ms,Maris:1999nt}.  Experimental data are taken 
from \cite{pdg} except for $f_{\eta_c}$ \cite{Edwards:2000bb}.}
\begin{center}
\begin{tabular}{l|ll|ll|ll|l|ll|ll}
\br
 $q$     & $M_{\pi_0}$ & $M_{\hbox{\scriptsize expt}}$
         & $f_{\pi_0}$ & $f_{\hbox{\scriptsize expt}}$
         & $M_{\pi_1}$ & $M_{\hbox{\scriptsize expt}}$
         & $f_{\pi_1}$
         & $M_{\rm V}$ & $M_{\hbox{\scriptsize expt}}$
         & $f_{\rm V}$ & $f_{\hbox{\scriptsize expt}}$ \\
\br
  $u,d$
        & 0.14 & 0.14   &  0.131 & 0.131
        & 1.10 & 1.3    & -0.002
        & 0.74 & 0.77   &  0.206 & 0.22  \\
  $s$
        & 0.70 & ---    &  0.182 & ---
        & 1.41 & 1.47   & -0.033
        & 1.08 & 1.02   &  0.257 & 0.23 \\
 $c$
        & 2.98 & 2.98   &  0.33 & 0.34
        & 3.45 & 3.65   & -0.15
        & 3.13 & 3.10   &  0.33 & 0.41 \\
\br
\end{tabular}
\end{center}
\end{table}
The main results for the pseudoscalar and vector meson masses and
leptonic decay constants are summarized in Table~\ref{Tab:excpion} and
illustrated in Fig.~\ref{Fig:mesons}.  Regarding the table one should
note that the rainbow-ladder truncation gives ``ideal'' flavor mixing
and the study uses $\hat m_u = \hat m_d \neq \hat m_s \neq \hat m_c$.
Thus the first row simultaneously describes four degenerate mesons for
each column, namely, a \{$u \bar d$, $u\bar u-d\bar d$,$d\bar u$\}
isotriplet and a $u\bar u+d\bar d$ isosinglet. The second row
describes $s\bar s$ mesons, and the third row $c\bar c$ mesons. For
vector mesons ideal mixing is a very good approximation, though this
is not the case for the pseudoscalar ground states.  However, the
experimental degeneracy of the $\pi(1300)$ and $\eta(1295)$
suggests~\cite{pdg} that ideal mixing is almost realized for the
excited pseudoscalar mesons.  This supports the interpretation of the
$\eta(1295)$ and $\eta(1470)$ as the radial excitations of the
$\eta(548)$ and $\eta^\prime(958)$, with quark content almost entirely
$u\bar u+d\bar d$ and $s\bar s$ respectively~\cite{Hoell:2004un}.
Figure~\ref{Fig:mesons} shows bound-state masses and leptonic decay 
constants for the ground and excited states of pseudoscalar mesons as well
as ground-state vector-meson masses as functions of the current-quark mass.
It illustrates that $m_{\pi_0}$ vanishes in the chiral limit, while
$m_{\rho}$ and $m_{\pi_1}$ do not; on the other hand, $f_{\pi_1}$
vanishes in the chiral limit, while $f_{\pi_0}$ does not.

Satisfaction of the axial-vector WTI can be checked by virtue of
Eq.~(\ref{eq:gmorgen}) for any pseudoscalar meson in
Table~\ref{Tab:excpion}.  For the ground and excited states the
inaccuracies are below 1\% and 5\%, respectively.  The latter are
larger because of: (1) the need to project out the ground state in
order to calculate the excited state, (2) the smallness of $f_{\pi_1}$
close to the chiral limit, and (3) the larger bound-state mass,
leading to a larger region of analytic continuation for the quark DSE
solution in the complex $p^2$ plane.

\section{Electromagnetic form factors}
Meson-meson-photon interactions can be described in generalized
impulse approximation by
\begin{equation}
  I^{abc}(P,Q,K) = \int\!\!\frac{d^4q}{(2\pi)^4} \,
    {\rm Tr}\big[ S^a(q) \, \Gamma^{a\bar{b}}(q,q';P)
        S^b(q')\,\Gamma^{b\bar{c}}(q',q'';Q) \, S^c(q'') \,
    \Gamma^{c\bar{a}}(q'',q;K) \big] \,,
\label{eq:generictri}
\end{equation}
where $q - q' = P$, $q' - q'' = Q$, $q'' - q = K$, and momentum
conservation dictates $P + Q + K = 0$.  In Eq.~(\ref{eq:generictri}),
$S^i$ is the dressed quark propagator with flavor index $i$, and
$\Gamma^{i\bar{j}}(k,k';P)$ stands for a generic vertex function with
incoming quark flavor $j$ and momentum $k'$, and outgoing quark flavor
$i$ and momentum $k$.  Depending on the specific process under
consideration, this vertex function could be a meson BSA or a
quark-photon vertex.

The quark-photon vertex, \mbox{$\Gamma_\mu(p_+,p_-;Q)$}, with $Q$ the
photon momentum and $p_\pm$ the quark momenta, is the solution of the
inhomogeneous BSE
\begin{equation}
 \Gamma_\mu(p_+,p_-;Q) = Z_2(\zeta) \, \gamma_\mu +
        \int\!\!\frac{d^4q}{(2\pi)^4} \, K(p,q;Q)
        \;S(q_+) \, \Gamma_\mu(q_+,q_-;Q) \, S(q_-)\, .
\label{verBSE}
\end{equation}
Solutions of the homogeneous version of Eq.~(\ref{verBSE})
define vector meson bound states at timelike photon momenta
\mbox{$Q^2=-M_{\rm V}^2$}.  It follows that $\Gamma_\mu(p_+,p_-;Q)$
has poles at these locations~\cite{emf} (see also the vector-meson
masses in Table~\ref{Tab:excpion}).

\subsection{Pion and kaon elastic form factors}
There are two diagrams that contribute to meson electromagnetic form
factors: one with the photon coupled to the quark and one with the
photon coupled to the antiquark respectively.  With photon momentum
$Q$, and incoming and outgoing meson momenta $P \mp Q/2$, we can
define a form factor for each of these diagrams~\cite{emf}
\begin{eqnarray}
    2\,P_\nu\,F_{a\bar{b};\bar{b}}(Q^2) &=&
            I_\nu^{a\bar{b};\bar{b}}(P-Q/2,Q,-(P+Q/2)) \,.
\label{eq:emff}
\end{eqnarray}
We work in the isospin symmetric limit, and thus \mbox{$F_{\pi}(Q^2) =
F_{u\bar{u};u}(Q^2)$}.  The $K^+$ and $K^0$ form factors are given by
\mbox{$F_{K^+}= \frac{2}{3}F_{u\bar{s};u} + \frac{1}{3}F_{u\bar s
;\bar s}$} and \mbox{$F_{K^0} = -\frac{1}{3}F_{d\bar{s};d} +
\frac{1}{3}F_{d\bar s ;\bar s}$}, respectively.

\begin{figure}[t]
\includegraphics[width=18pc]{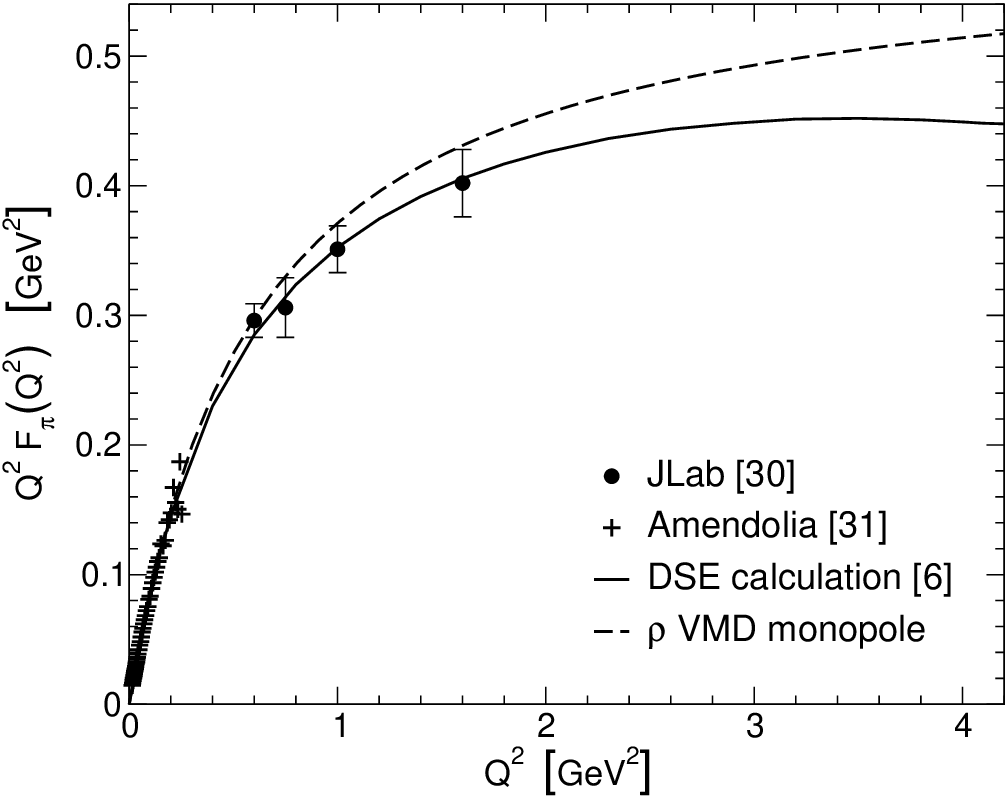}\hspace{1pc}%
\includegraphics[width=18pc]{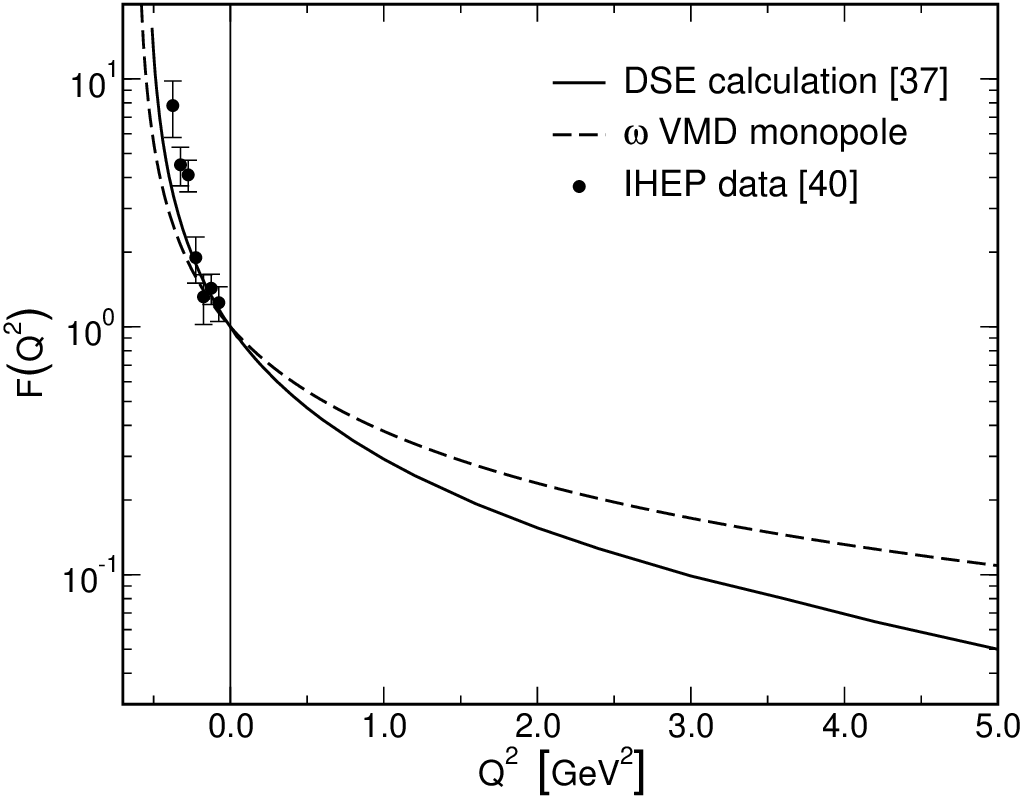}
\caption{\label{Fig:emff} 
Left: The pion form factor $Q^2 F(Q^2)$, compared to the 
experimental results from Refs.~\cite{Volmer:2000ek,Amendolia:1986wj}.
Right: The transition form factor $F_{\omega\pi\gamma}(Q^2)$ together
with experimental data~\cite{Dzhelyadin:1980tj}.}
\end{figure}
Our result for $Q^2 F_\pi$ are shown in the left panel of
Fig.~\ref{Fig:emff}.  In the timelike region, and in the spacelike
region up to about $Q^2 = 2\,{\rm GeV}^2$, $F_\pi$ can be described
very well by a monopole with our calculated $\rho$-mass,
$m_\rho=742\,{\rm MeV}$ (note that our calculated $\rho$-mass is
slightly below the experimental value).  Above this value, our curve
starts to deviate more and more from this naive vector meson dominance
(VMD) monopole.  Our result is in excellent agreement with the most
recent JLab data~\cite{Volmer:2000ek}; it would be very interesting to
compare with future JLab data in the 3 to 5 GeV$^2$ range, where we
expect to see a significant deviation from the naive monopole
behavior.  The pQCD prediction is significantly smaller than our
results at 4 GeV$^2$; we anticipate that true perturbative behavior
sets in somewhere between 10 and 20 GeV$^2$~\cite{Maris:1998hc}.

Recent lattice simulation also indicate that in the region between 0
and 2 GeV$^2$ the pion form factor can be represented by a VMD-like
monopole~\cite{vanderHeide:2003kh,Bonnet:2004fr}.  This VMD-like
behavior of the pion form factor in the spacelike region appears to be
valid almost independent of the pion mass, both in our calculations
and in the lattice simulations, though a monopole fit results in a
VMD-mass which is slightly less than the actual vector meson mass.
Current lattice simulations are not accurate enough, nor do they
extend to large enough values of $Q^2$, to detect a significant
deviation from VMD-like behavior above 2 GeV$^2$.

\begin{table}[b]
\caption{\label{Tab:radiiradec}
Overview of our results for the pseudoscalar-meson charge radii
squared, all in ${\rm fm}^2$, with an estimated combined numerical
error of less than $0.01~{\rm fm}^2$.  Also included are our results
for the vector meson radiative decays: $\Gamma_{V\to P\gamma}$ in keV
and $g/m$ in GeV$^{-1}$.}
\begin{center}
\begin{tabular}{l|lll|ll|ll|ll|ll}
\br
& $r^2_\pi$  & $r^2_{K^+}$ & $r^2_{K^0}$
& $\Gamma_{\rho^\pm}$ & $g/m$ 
& $\Gamma_{\omega}$   & $g/m$ 
& $\Gamma_{K^{\star\pm}}$ & $g/m$ 
& $\Gamma_{K^{\star 0}}$  & $g/m$ \\
\br
calc. & 0.44 & 0.38 & -0.085
           & 53 & 0.69 & 479 & 2.07 & 90 & 0.99 & 130 & 1.19 \\
expt. & 0.44 & 0.34(5)& -0.052(26)
           & 68 & 0.74 & 757 & 2.38 & 50 & 0.84 & 116 & 1.27 \\
\br
\end{tabular}
\end{center}
\end{table}
Also our results~\cite{emf} for $F_K$ agree quite well with the
available experimental data, as do both the neutral and the charged
kaon charge radius~\cite{Amendolia:1986ui,Molzon:1978py}, see
Table~\ref{Tab:radiiradec}.  It should be noted here that the $K^0$
form factor is obtained by taking the difference of two numbers,
$F_{d\bar{s};d}(Q^2)$ and $F_{d\bar s ;\bar s}(Q^2)$, which are both
close to one for $Q^2$ near zero.  It is therefore much more sensitive
to details of the model: both kaon loops and a flavor dependence of
the (effective) quark-quark interaction will have a significantly
bigger effect on the $K^0$ form factor than on the $K^+$ form factor.
This could explain why it does not agree as well with experiment as
our other results.  In this respect one should note that the absolute
deviation between our calculated charge radius and the experimental
charge radius is similar for the charged and neutral kaon.

\subsection{Radiative vector-meson decays}
We can describe the radiative decay of the vector mesons using the
same loop integral, Eq.~(\ref{eq:generictri}), this time with one
vector meson BSA, one pseudoscalar BSA, and one
$q\bar{q}\gamma$-vertex~\cite{Maris:2002mz}.  The generic structure of
a vector-pseudoscalar-photon vertex is
\begin{eqnarray}
 I^{a\bar{b};a}_{\mu \nu}(P,Q,-(P+Q)) & = &
  \epsilon_{\mu\nu\alpha\beta}  P_\alpha Q_\beta  \, 
  f_{a\bar{b};a}(Q^2) \,,
\end{eqnarray}
where $P$ is the vector momentum and $Q$ the photon momentum.  The
on-shell value gives us the coupling constant, and can be used to
calculate the partial decay width of the vector mesons.  For virtual
photons, we can define a transition form factor $F_{VP\gamma}(Q^2)$,
normalized to 1 at $Q^2 = 0$, which can be used in estimating
meson-exchange contributions to hadronic
processes~\cite{VanOrden:1995eg,Tandy:1997qf}.

In the isospin limit, both the $\rho^0\,\pi^0\,\gamma$ and
$\rho^\pm\,\pi^\pm\,\gamma$ vertices are given by
\begin{eqnarray}
\label{eq:rpgver}
  \frac{g_{\rho\pi\gamma}}{m_\rho} \; \epsilon_{\mu\nu\alpha\beta}
    P_\alpha Q_\beta  \, F_{\rho\pi\gamma}(Q^2)  &=&
 \case{1}{3} \, I^{u\bar{u};u}_{\mu \nu}(P,Q,-(P+Q)) \,.
\end{eqnarray}
The $\omega\,\pi\,\gamma$ vertex is a factor of 3 larger, due to the
difference in isospin factors; however, the form factor
$F_{VP\gamma}(Q^2)$ is the same for $\rho\,\pi\,\gamma$ and
$\omega\,\pi\,\gamma$.  In contrast to the elastic form factors, this
transition form factor falls off significantly faster than a VMD-like
monopole, as can be seen in the right panel of Fig.~\ref{Fig:emff}.
Only in the timelike region, near the vector meson pole, do we see a
true VMD-like behavior.

As Eq.~(\ref{eq:rpgver}) shows, it is $g_{VP\gamma}/m_V$ that is the
natural outcome of our calculations.  Therefore, it is this
combination that we give in Table~\ref{Tab:radiiradec}, together with
the corresponding partial decay widths~\cite{Maris:2002mz}.  As
anticipated, the partial decay width $\omega \to \pi\gamma$ is indeed
(approximately) nine times larger than the $\omega \to \pi\gamma$
partial decay width.  Note that part of the difference between the
experimental and calculated decay width comes from the phase space
factor because our calculated vector meson masses deviate up to 5\%
from the physical masses.

For the $K^\star K \gamma$ decays (and corresponding form
factors)~\cite{Maris:2002na} the situation is more complicated owing to
the interference of the diagrams with the photon coupled to the
$s$-quark and to the $u$- or $d$-quark.  In the SU(3) flavor limit,
the charged $K^\star \, K \, \gamma$ vertex becomes equal to the
$\rho\,\pi\,\gamma$ vertex, whereas the neutral $K^\star \, K \,
\gamma$ vertex is twice as large in magnitude
\begin{eqnarray}
 \left(\frac{g_{K^\star K \gamma}}{m_{K^\star}}\right)^+ \;
  F^{+}_{K^\star K \gamma}(Q^2) &=&
           \case{2}{3} f_{u\bar{s};u}(Q^2)
                        - \case{1}{3} f_{u\bar{s};\bar{s}}(Q^2) \,,
\\
 \left(\frac{g_{K^\star K \gamma}}{m_{K^\star}}\right)^0 \;
  F^{0}_{K^\star K \gamma}(Q^2) &=&
           -\case{1}{3} f_{d\bar{s};d}(Q^2)
                        - \case{1}{3} f_{u\bar{s};\bar{s}}(Q^2) \,.
\end{eqnarray}
In Table~\ref{Tab:radiiradec} we see indeed that the partial decay
width of the neutral $K^\star \to K \gamma$ is larger than that of the
charged $K^\star \to K \gamma$, though not by a factor of four as it
would in the SU(3) flavor limit.  Furthermore, we see that the
deviation between our calculation and experiment is largest for the
charged $K^\star \to K \gamma$ decay.  Again, this can be understood
since this decay is sensitive to the difference between the impulse
diagrams, and therefore more sensitive to details of the model and its
omissions.

\ack
We would like to thank Craig Roberts and Peter Tandy for useful
discussions.  This work was supported by the Austrian Research
Foundation \textit{FWF, Erwin-Schr\"odinger-Stipendium} no.\
J2233-N08, the Department of Energy, Office of Nuclear Physics,
contract no.\ W-31-109-ENG-38, and benefited from the ANL Computing
Resource Center's facilities; part of the computations were performed
on the National Science Foundation Terascale Computing System at the
Pittsburgh Supercomputing Center.


\end{document}